\documentclass[aps,prl,10pt,twocolumn,amsmath,citeautoscript,groupedaddress,superscriptaddress]{revtex4-2}

\usepackage[colorlinks,urlcolor=blue,citecolor=blue]{hyperref}
\usepackage{color}
\usepackage{graphicx}

\begin{document}

\title{Anomalous stopping and charge transfer in proton-irradiated graphene}

\author{Alina Kononov}
\affiliation{Department of Physics, University of Illinois at Urbana-Champaign, Urbana, IL 61801, USA}
\author{Andr\'{e} Schleife}
\email{schleife@illinois.edu}
\affiliation{Department of Materials Science and Engineering, University of Illinois at Urbana-Champaign, Urbana, IL 61801, USA}
\affiliation{Materials Research Laboratory, University of Illinois at Urbana-Champaign, Urbana, IL 61801, USA}
\affiliation{National Center for Supercomputing Applications, University of Illinois at Urbana-Champaign, Urbana, IL 61801, USA}

\date{\today}

\begin{abstract}
We use first-principles calculations to uncover and explain a new type of anomalous low-velocity stopping effect in proton-irradiated graphene.
We attribute a shoulder feature that occurs exclusively for channeling protons to enhanced electron capture from $\sigma$+$\pi$ orbitals.
Our analysis of electron emission indicates that backward emission is more sensitive to proton trajectory than forward emission and could thus produce higher contrast images in ion microscopy.
For slow protons, we observe a steep drop in emission, consistent with predictions from analytical models.
\end{abstract}

\maketitle

\textit{Introduction} ---
Graphene, an exceptionally strong and intrinsically semimetallic sheet of carbon atoms arranged in a honeycomb lattice \cite{novoselov:2004,novoselov:2005,geim:2007,castro_neto:2009}, was the first 2D material ever fabricated \cite{novoselov:2004} and remains promising as a conductive layer in novel electronics based on 2D heterostructures \cite{geim:2013}.
It has attracted enormous interest due to its potential for transistors \cite{novoselov:2004,lin:2009,schwierz:2010,bertolazzi:2013}, solar cells \cite{sun:2011}, detectors \cite{allegrini:2014,skoblin:2018}, plasmonics \cite{grigorenko:2012}, qubits \cite{wang:2019}, and other devices \cite{sun:2011,surwade:2015}.
However, graphene's sensitivity to defects and nanostructural features, which can degrade performance when undesirable \cite{tapaszto:2008,vicarelli:2015} or enable an application when intentional \cite{zhao:2012,bangert:2013,bell:2009,kotakoski:2015,li:2017:ions2D}, poses a major obstacle for scalable fabrication of graphene-based devices.

Consequently, precise techniques for characterization, patterning, and defect engineering of 2D materials including graphene are necessary to realize their incredible potential.
Focused ion beams are promising in this context, with several empirical demonstrations of their capability to image \cite{fox:2013,zhou:2014}, characterize \cite{guo:2014}, create point defects in \cite{zhao:2012,bangert:2013,iberi:2016}, otherwise alter atomic structure of \cite{bell:2009,ritter:2013,hemamouche:2014,kotakoski:2015,wilhelm:2015,fox:2015,iberi:2015, naitou:2015}, and tune mechanical \cite{iberi:2016} and electronic \cite{fox:2015} properties of 2D materials.
However, ion beam parameters must be specially tuned for 2D materials because they exhibit a highly \emph{pre-equilibrium} response to ion irradiation which differs from bulk and thin films:
Highly-charged ions partially neutralize \cite{lee:2020,wilhelm:2018} and reach an equilibrium charge state only after traversing $\sim$10\,nm of material \cite{schenkel:1997,hattass:1999}, leading to deviations from bulk behavior for atomically thin systems \cite{wilhelm:2014}.
Even in the case of proton irradiation, surface plasmons are predicted to enhance energy deposition \cite{kononov:2020}, and the radically different plasmonic properties of 2D materials \cite{eberlein:2008,kinyanjui:2012} should further influence charge and energy transfer processes upon ion impact.

Despite the practical importance of ion beam techniques, a detailed understanding of the unique physics occurring during an energetic ion's traversal of a 2D material remains elusive.
While experimental work routinely identifies intriguing results of applying specific beam parameters to particular materials, direct observation of the underlying mechanisms is limited by spatial and temporal resolutions of measurement techniques.
Accurate first-principles calculations offer a promising alternative, enabling extremely detailed simulations of fs-scale dynamics after a single ion impact.
Indeed, numerous studies \cite{schleife:2015,maliyov:2018,zeb:2012,quashie:2016,quashie:2018, yost:2016,yost:2017,lee:2018,lee:2020, pruneda:2007,zeb:2013,mao:2014,li:2017, ullah:2015,lim:2016} have demonstrated the predictive power of such calculations for \emph{bulk} materials under ion irradiation, showing, for instance, that directional bonding in semiconductors and insulators makes electronic stopping sensitive to ion trajectory even for slow ions, when core electrons are negligible \cite{ullah:2015,lee:2018,lee:2020}.

Additionally, computational studies \cite{pruneda:2007,zeb:2013,mao:2014,li:2017, ullah:2015,lim:2016} have elucidated experimentally observed \cite{serkovic:2007,markin:2009:insulators,roth:2013} threshold effects in low-velocity stopping power, where exciting electrons across a band gap requires a minimum projectile velocity.
Similar effects can be expected for electron emission, where transfering sufficient energy for electrons to overcome the work function would require a minimum projectile velocity.
Although experiments support their existence \cite{baragiola:1979,markin:2008}, to date no first-principles study has examined such threshold emission effects.
Furthermore, graphene presents a highly interesting case with
directional bonding but no band gap.
While previous work \cite{miyamoto:2008,zhang:2012,bubin:2012,ojanpera:2014,zhao:2015,gruber:2016,he:2021} has simulated ion-irradiated 2D materials, none has established the connection between energy transfer, charge capture, and electron emission, or even separately analyzed emissions from both sides of the material.

In this Letter, we present a comprehensive, first-principles computational study of proton-irradiated graphene and discover a new type of anomalous effect in energy deposition by slow ions.
This effect only occurs along some proton trajectories and is accompanied by enhanced projectile electron capture.
Furthermore, calculations of emitted electron yields from both sides of the material lead to predictions relevant to ion beam microscopy and the first first-principles study of threshold electron emission.

\textit{Methods} ---
We performed real-time time-dependent density functional theory (TDDFT) \cite{runge:1984,marques:2004,marques:2006,ullrich:2011,ullrich:2014} simulations of the excited electron dynamics in proton-irradiated monolayer graphene using the Qbox/Qb@ll code \cite{gygi:2008,schleife:2012,schleife:2014,draeger:2017,draeger:2018}.
Converged ground-state single-particle Kohn-Sham (KS) states from DFT \cite{dreizler:1990} served as initial condition for real-time propagation of the time-dependent KS equations, 
\begin{equation}
i\frac{\partial}{\partial t} \phi_j(\mathbf{r},t) = \left(-\frac{\nabla^2}{2} + V_{\mathrm{KS}}[n](\mathbf{r},t)\right)\phi_j(\mathbf{r},t).
\label{eq:TDKS}
\end{equation}
Here, $\phi_j$ are single-particle KS orbitals evolving in a time-dependent effective potential $V_{\mathrm{KS}}$ which is a functional of the electron density
\begin{equation}
n(\mathbf{r},t) = \sum_j f_j |\phi_j(\mathbf{r},t)|^2
\label{eq:density}
\end{equation}
and $f_j$ are orbital occupations.

KS states are represented in a plane-wave basis with a cutoff energy of 100\,Ry, exchange and correlation (XC) is treated with the adiabatic local density approximation \cite{zangwill:1980,zangwill:1981}, and the electron-ion interaction is described using HSCV pseudopotentials \cite{vanderbilt:1985}.
Atomic forces on all carbon atoms in the initial structure were relaxed to less than 2\,meV/{\AA}.
Large simulation cells with 112 carbon atoms and 150\,a$_0$ of vacuum were needed to converge electron emission, allowing reciprocal-space sampling using only the $\Gamma$ point.
We address convergence with respect to cutoff energy and vacuum size in Figs.\ S1 and S2 of the Supplemental Material (SM).
To help interpret our TDDFT results, a projected density of states (DOS) for ground-state graphene was calculated using VASP \cite{kresse:1996,kresse:1999} with a 4-atom supercell, $64\times 64\times 1$ $\Gamma$-centered Monkhorst-Pack \cite{monkhorst:1976} $\mathbf{k}$ point mesh, and the same vacuum length, cutoff energy, and XC approximation as above;
the obtained DOS is similar to previously reported results \cite{trickey:1992,ooi:2006,klintenberg:2009}.

\begin{figure}
\centering
\raisebox{0.48in}{\includegraphics{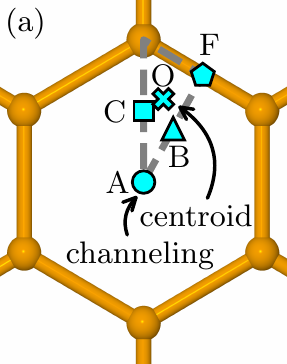}}\hfill
\includegraphics{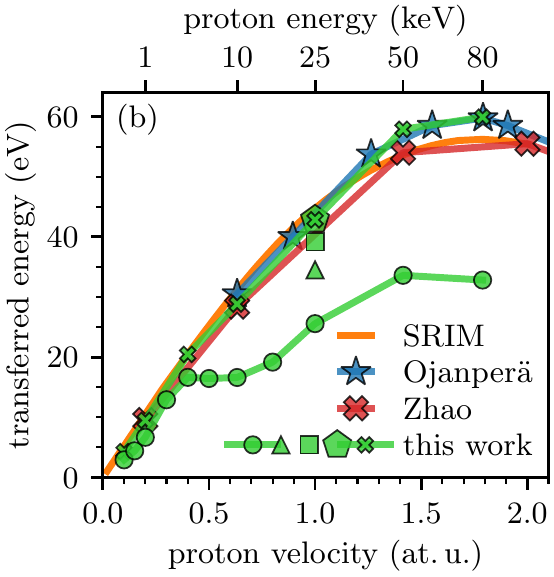}
\caption{\label{fig:Etransfer}(Color online.)
(a) Proton impact points (cyan) in monolayer graphene (orange) investigated here.
The projectile always travels normal to the graphene plane.
(b) Total energy deposited in graphene by energetic protons.
Results from this work are compared to SRIM \cite{ziegler:2010} and previous TDDFT results \cite{ojanpera:2014,zhao:2015}.
Symbols correspond to the different impact points illustrated in (a).
}
\end{figure}

We begin real-time propagation with the proton 25\,a$_0$ away from graphene.
It then approaches and traverses the material at a constant velocity along a normal trajectory (see Fig.\ \ref{fig:Etransfer}a).
Carbon nuclei are held fixed because the few-fs time-scale of the simulations is too short for them to move appreciably.
For numerical integration of the time-dependent KS equations, we use the enforced time reversal symmetry (ETRS) method \cite{castro:2004,draeger:2017} with a time step of 1.0 atto-second.
This choice has been shown to evolve similarly large systems with exceptional accuracy \cite{kang:2019, kononov:2020}.

The Hellmann-Feynman force on the projectile gives the instantaneous electronic stopping power, or rate of energy transfer from the proton to the material.
To obtain the total energy transferred, the instantaneous stopping data (see Fig.\ S4 in the SM) is integrated over a layer thickness taken as the interlayer separation in graphite, though we find very similar trends for other possible choices such as the dielectric thickness reported in Ref.\ \onlinecite{rickhaus:2020} (see Fig.\ S5 in the SM).

Upon proton impact, electrons are emitted from both sides of the graphene.
Entrance- and exit-side emitted electron yields are determined by integrating the time-dependent electron density from the TDDFT simulations over the corresponding vacuum regions (see details in \mbox{Fig.\ S3} of the SM) and subtracting electrons captured by the projectile, as determined by fits to analytical orbitals as described in Ref.\ \onlinecite{kononov:2020}.
All inputs and outputs from our simulations are available at the Materials Data Facility \cite{MDF,data}.

\textit{Results and Discussion} ---
While proton-irradiated graphene has been simulated before \cite{bubin:2012,ojanpera:2014,zhao:2015}, including some exploration of different impact points \cite{bubin:2012,ojanpera:2014}, a comprehensive understanding of the trajectory dependence of energy deposition remains absent.
Our analysis for different impact points (see \mbox{Fig.\ \ref{fig:Etransfer}a}) allows us to interpret the underlying physics in terms of spatial inhomogeneity of the electron density arising from chemical bonding in graphene.
For the centroid impact point, which is often considered a good approximation to an ensemble average over all trajectories \cite{ojanpera:2014,yost:2016,maliyov:2018}, our results for energy transferred from proton to graphene agree well with earlier TDDFT \cite{ojanpera:2014,zhao:2015} and empirical \cite{ziegler:2010} data for per-layer electronic stopping power of hydrogen in graphite (see Fig.\ \ref{fig:Etransfer}b).
We discuss potential sources of minor discrepancies among computed energy transfer results in the SM.

Interestingly, we find that energy transfer is quite sensitive to the impact point:
We compute considerably lower electronic stopping for the channeling than for the centroid trajectory at velocities above 0.4\,at.\,u.\ (see Fig.\ \ref{fig:Etransfer}b).
This contrasts with behavior in bulk metals, where valence electrons are largely delocalized and stopping power depends strongly on the projectile's trajectory only for velocities sufficiently high to excite localized core electrons \cite{schleife:2015,maliyov:2018}.
However, our finding confirms earlier reports of higher energy deposition when the proton passes through a C\,\---\,C bond than through the center of the hexagon \cite{bubin:2012,ojanpera:2014}.
Trajectory-dependent stopping power has also been reported in bulk semiconductors such as Si \cite{lee:2020}, Ge \cite{ullah:2015}, and phosphide-based compounds \cite{lee:2018}.
In these materials, directional bonding leads to differences in the electron density with which the projectile interacts when moving along different channels, causing strong trajectory dependence of electronic stopping.

\begin{figure}
\centering
\includegraphics{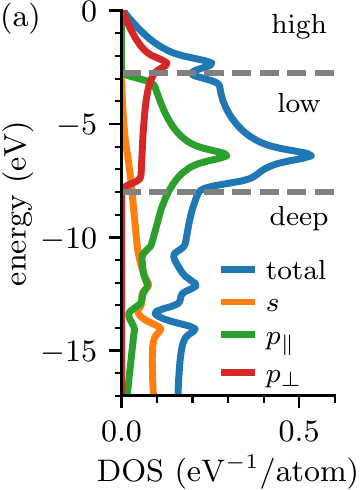} \hfill
\raisebox{0.05in}{\includegraphics{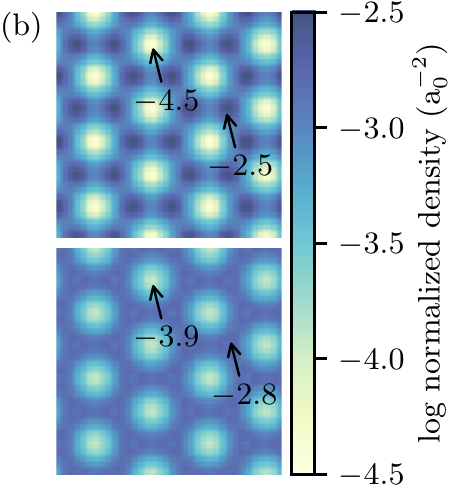}}
\caption{\label{fig:GS_den_profile}(Color online.)
(a) Projected density of states of ground-state graphene with valence band divisions indicated by horizontal dashed lines (see text).
Energies are referenced to the valence band maximum.
(b) Ground-state electron density contributed by the high-energy (top) and low-energy (bottom) graphene valence bands.
Electron density has been integrated along the direction normal to graphene and is normalized by the number of electrons within each set of bands.
Annotations indicate minimum and maximum values.
}
\end{figure}

Notably, the above studies finding channel-dependent stopping power in bulk materials nonetheless observed the same qualitative behavior as a function of projectile velocity across different channels, i.e., a featureless rise \cite{ullah:2015} toward a single peak \cite{lee:2018,lee:2020} positioned near the experimental maximum even while the peak height depends on projectile trajectory.
Conversely, we found that energy transferred along a channeling trajectory in graphene exhibits an unusual, bimodal velocity dependence featuring a shoulder between 0.4 and 0.7\,at.\,u.\ of velocity (see Fig.\ \ref{fig:Etransfer}).
Similar features appeared in Bubin \emph{et al.}'s results for proton-irradiated graphene fragments \cite{bubin:2012}, but no explanation was offered.
Deviations from linear low-velocity stopping are typically attributed to band structure effects: In insulators \cite{serkovic:2007,markin:2009:insulators,pruneda:2007,zeb:2013,mao:2014,li:2017} and semiconductors \cite{roth:2013,ullah:2015,lim:2016}, slow projectiles cannot excite electrons across the band gap, and in metals \cite{figueroa:2007,markin:2008,cantero:2009,markin:2009:metals,zeb:2012,quashie:2016,quashie:2018}, slow projectiles cannot excite electrons from deeper valence bands.
However, to our knowledge, no study has explained the anomalous trajectory dependence of stopping for semimetallic graphene.

To explain this behavior, we examine both the electronic structure and the spatial distributions of partial electron densities of graphene in Fig.\ \ref{fig:GS_den_profile}.
Our analysis reveals that high-energy states lying less than 2.75\,eV below the valence band maximum (VBM), commonly known as $\pi$ electrons, are more localized in-plane around the carbon atoms than lower-energy states (between 2.75 and 8\,eV below the VBM), consisting of both $\pi$ and $\sigma$ orbitals.
This spatial inhomogenity and spectral separation suggests that high-energy states could be more difficult for a channeling proton to excite, and below we explore how these observations affect velocity-dependent stopping.

\begin{figure}
\centering
\includegraphics{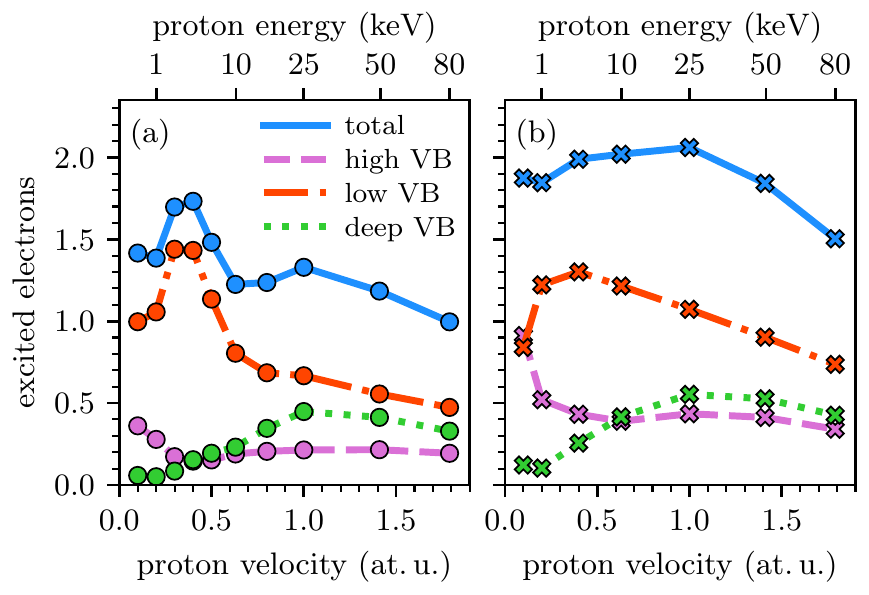}
\caption{\label{fig:eexcited}
(Color online.)
Number of electrons excited after a proton traverses graphene (a) along the channeling trajectory and (b) along the centroid trajectory.
The total number of excited electrons is decomposed into contributions from the different sets of valence bands illustrated in Fig.\ \ref{fig:GS_den_profile}a.
}
\end{figure}

Specifically, to investigate the source of the shoulder-like feature in the stopping power of channeling protons, we examined the occupation of the original, ground-state orbitals after irradiation and the extent to which the proton excites electrons from different bands.
We obtained the excited electron contribution $n_j(t)$ from a ground-state KS orbital $|\phi_j^{(\mathrm{GS})}\rangle$ by summing over projections of the time-dependent KS orbitals $|\phi_\ell(t) \rangle$,
\begin{equation}
\label{eq:proj}
n_j(t) = 2\left(1 -\sum_\ell \left|\langle \phi_\ell(t) | \phi_j^{(\mathrm{GS})}\rangle\right|^2 \right),
\end{equation}
where the factor of 2 accounts for spin degeneracy.
Subsequently, the $n_j(t)$ were summed over states within the particular energy bands illustrated in Fig.\ \ref{fig:GS_den_profile}a; the results become constant over time (see Fig.\ S6 in the SM), and we report the number of electrons excited from each band at the end of each simulation in Fig.\ \ref{fig:eexcited}.

Figure \ref{fig:eexcited} shows that the shoulder in the energy transfer for channeling protons is associated with a drop in the total number of electrons excited.
Analysis of the contributions of different valence bands provides deeper insight:
While a channeling proton indeed excites high-energy valence electrons at lower rates than a proton traversing the centroid trajectory (as suggested by the electron density distributions in Fig.\ \ref{fig:GS_den_profile}), this effect remains nearly constant throughout the entire velocity range.
Unexpectedly, the most striking difference between the excitations induced along different proton trajectories occurs in the low-energy valence bands,
where the number of electrons excited by the channeling proton exhibits a much more pronounced maximum near velocities of 0.3\,--\,0.4\,at.\,u.
We thus conclude that the features in the total number of excited electrons and in electronic stopping near those velocities mainly originate from the low-energy valence bands.
Finally, the extent of deep valence band excitation is not sensitive to projectile trajectory in this velocity range.

\begin{figure}
\centering
\includegraphics{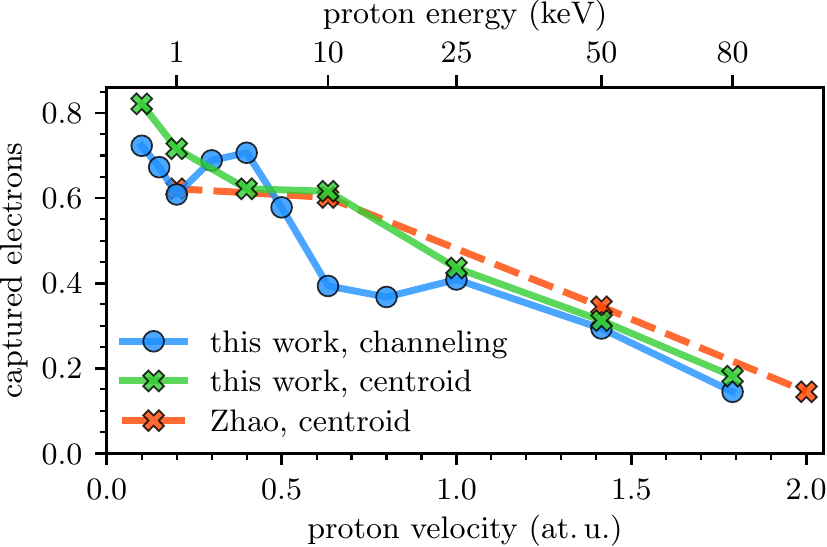}
\caption{\label{fig:Qtransfer}(Color online.) 
Electrons captured by the proton after transmission through graphene along channeling and centroid trajectories.
Results from this work are compared to previous TDDFT results \cite{zhao:2015}.
}
\end{figure}

Analyzing the charge captured by the projectile after transmission through graphene offers additional insight into the physics of trajectory-dependent stopping.
For the centroid trajectory, our results for electron capture agree well with previous calculations \cite{zhao:2015} and show the same trend of decreasing capture for faster protons (see Fig.\ \ref{fig:Qtransfer}) reported for proton-irradiated aluminum \cite{kononov:2020}.
However, for channeling protons, the enhanced excitation at velocities of 0.3\,--\,0.4\,at.\,u.\ (see Fig.\ \ref{fig:eexcited}) is accompanied by a similar feature in the number of electrons captured by the projectile (see Fig.\ \ref{fig:Qtransfer}).
This suggests that resonant electron capture from the low-energy valence bands may be responsible for the shoulder in energy transfer for channeling protons.
Experiments such as Ref.\ \onlinecite{wilhelm:2014} which simultaneously measure projectile energy loss and charge state could potentially confirm this prediction.

\begin{figure}
\centering
\includegraphics[width=\columnwidth]{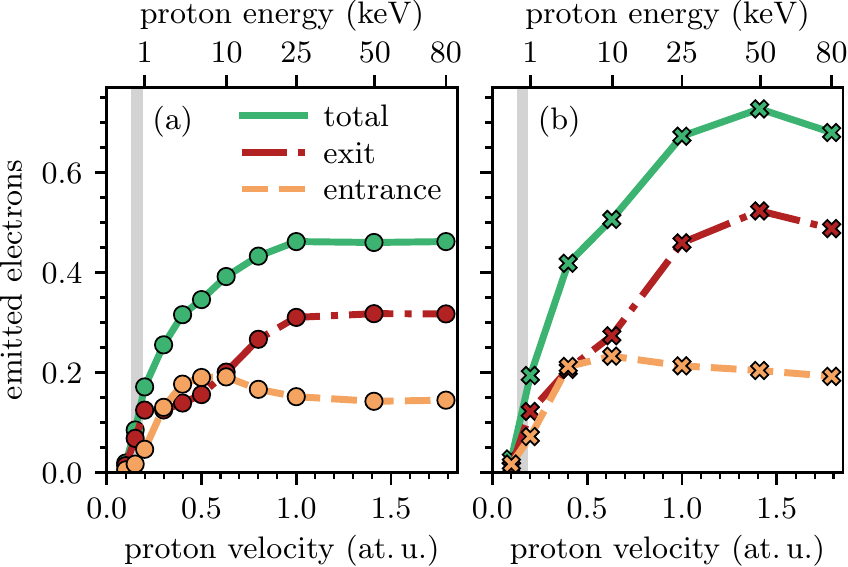}
\caption{\label{fig:emission}
(Color online.)
Electrons emitted after a proton traverses graphene along the (a) channeling and (b) centroid trajectory.
The total number of emitted electrons is decomposed into exit-side and entrance-side emission.
The gray bar indicates the range of threshold velocities predicted by different analytic models.
}
\end{figure}

While our above analysis of energy deposition can help explain eventual radiation damage, ion-induced electron emission has additional implications for imaging \cite{ward:2006,notte:2007,hlawacek:2014, fox:2013,zhou:2014}.
Here, we calculate emitted electron yields for the side from which the projectile approaches before impact (entrance-side) and the side from which it emerges (exit-side).
Interestingly, the exit-side and entrance-side emissions shown in Fig.\ \ref{fig:emission} exhibit very different velocity dependences:
Maximum entrance-side emission occurs at a velocity of 0.63\,at.\,u., while the peak in exit-side emission occurs much closer to the stopping maximum, at a velocity of 1.41\,at.\,u.
Furthermore, the exit-side emission is more sensitive to proton trajectory, particularly near the emission maximum, where the centroid trajectory yields 65\,\% more electrons than the channeling trajectory.
These findings indicate that detecting exit-side electron emission after irradiation by 50\,keV protons may provide high contrast images of graphene samples, especially since protons at this energy are not expected to damage the atomic structure \cite{bubin:2012}.

Finally, our first-principles results also describe behavior near the theoretically expected but experimentally elusive kinetic emission threshold for slow projectiles.
We observe a steep drop in electron emission, with less than 0.02 electrons emitted for a proton with 0.1\,at.\,u.\ of velocity (see \mbox{Fig.\ \ref{fig:emission}}).
This is consistent with the threshold velocities $v_{\mathrm{th}}=0.13$\,--\,0.19\,at.\,u.\ predicted by
several analytic models \cite{baragiola:1979,lim:2016,quashie:2016,correa:2018} (see details in Eqs.\ (S1)\,\---\,(S3) of the SM).
However, these estimates for $v_{\mathrm{th}}$ can be understood to predict the velocity below which the projectile cannot excite electrons \textit{directly} into the continuum, though it could still, for instance, excite electrons into conduction bands and subsequently promote them into the continuum.
Thus, the low but non-zero emitted electron yields that we calculate for a proton velocity of 0.1\,at.\,u.\ may represent nonlinear or sub-threshold emission, which has been indirectly measured \cite{lakits:1990}.

We note that in addition to such sub-threshold effects, numerical uncertainties also limit the conclusiveness of our emission predictions for proton velocities below 0.2\,at.\,u.
This includes the sensitivity of low emission data to the precise definition of the material-vacuum boundary, since accurately distinguishing emitted electrons with extremely low kinetic energy $E$ from an excess electron density on the graphene surface would require very long simulation times scaling as $1/\sqrt{E}$.
We do not expect recapturing of electrons by graphene to introduce uncertainties, since this process does not occur during our few-fs simulations (see Fig.\ S7 of the SM) and the positive charge induced in the graphene spreads out too quickly to attract escaping electrons strongly.

\textit{Conclusions} ---
We found an anomaly in low-velocity energy deposition by channeling protons in graphene which does not occur for protons traversing a centroid trajectory.
This behavior is accompanied by pronounced features in the numbers of electrons excited from lower-energy graphene valence states and those captured by the proton.
Thus, we propose that this anomalous behavior is caused by resonant projectile charge capture from the lower-energy valence bands.
Since a projectile's charge equilibrates after traversing a few nanometers of
material, this effect may be specific to few-layer materials.

We also characterized electron emission from proton-irradiated graphene and found that exit-side emission is more sensitive to proton trajectory than entrance-side emission, making it a strong candidate for high-contrast imaging techniques.
A steep drop in emitted electron yields for slow protons is consistent with threshold velocities predicted by analytical models.
The quantitative predictions of our first-principles results open a new avenue for optimizing parameters for ion beam techniques and studying the fundamental physics of threshold emission effects in real materials.

\begin{acknowledgments}
This material is based upon work supported by the National Science Foundation under Grant No.\ OAC-1740219.
Support from the IAEA F11020 CRP ``Ion Beam Induced Spatio-temporal Structural Evolution of Materials: Accelerators for a New Technology Era" is gratefully acknowledged.
This research is part of the Blue Waters sustained-petascale computing project, which is supported by the National Science Foundation (awards OCI-0725070 and ACI-1238993) and the state of Illinois.
Blue Waters is a joint effort of the University of Illinois at Urbana-Champaign and its National Center for Supercomputing Applications.
This work made use of the Illinois Campus Cluster, a computing resource that is operated by the Illinois Campus Cluster Program (ICCP) in conjunction with the National Center for Supercomputing Applications (NCSA) and which is supported by funds from the University of Illinois at Urbana-Champaign.
\end{acknowledgments}

\clearpage




\title{Supplemental Material: Anomalous stopping and charge transfer in proton-irradiated graphene}
\author{Alina Kononov}
\author{Andr\'{e} Schleife}
\email{schleife@illinois.edu}
\affiliation{University of Illinois at Urbana-Champaign}
\date{\today}

\maketitle

\renewcommand\thefigure{S\arabic{figure}}
\renewcommand{\theHfigure}{S\thefigure}
\renewcommand\theequation{S\arabic{equation}}
\renewcommand{\theHequation}{S\theequation}
\renewcommand{\thetable}{S\Roman{table}}
\renewcommand{\theHtable}{S\thetable}
\setcounter{figure}{0}
\setcounter{equation}{0}

\section{Convergence of Cutoff Energy}
\label{sec:converge_ecut}

\begin{figure}[h]
\centering
\includegraphics{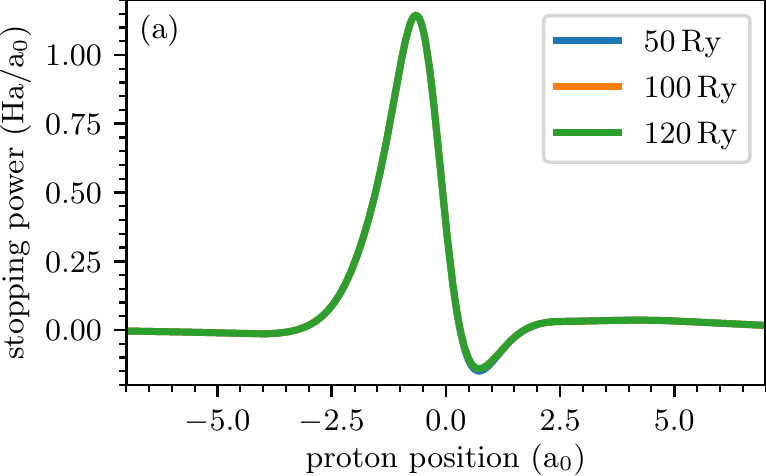}\\
\vspace{0.1in}
\includegraphics{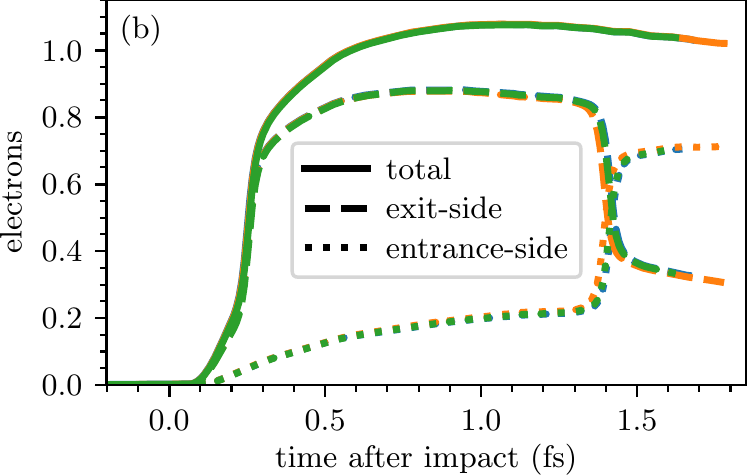}
\caption{(a) Instantaneous stopping power and (b) the number of electrons found in the vacuum (decomposed into exit-side and entrance-side regions) as a proton with 1\,at.\,u.\ of velocity impacts graphene at the midpoint of a C\,\---\,C bond do not change noticeably as the plane-wave cutoff energy is increased from 50\,Ry (blue) to 100\,Ry (orange) or 120\,Ry (green).
The stopping power (total electron) curves differ by less than 0.01\,Ha/a$_0$ (0.05 electrons).
}
\label{fig:ecut_convergence}
\end{figure}

We tested convergence with respect to the size of the plane wave basis by comparing results computed for a proton with 1\,at.\,u.\ of velocity using cutoff energies of 50\,Ry, 100\,Ry, and 120\,Ry.
To reduce computational cost, these tests were performed on a smaller supercell containing only 60 carbon atoms and 100\,a$_0$ of vacuum. 
These cutoff energies produce essentially identical results (see \mbox{Fig.\ \ref{fig:ecut_convergence}}), with differences of less than 1\% in the total energy deposited and the total charge induced in the graphene.
Therefore, we conclude that a cutoff energy of 100\,Ry is adequate for converged simulations of energy and charge dynamics in ion-irradiated graphene.

\section{Convergence of Vacuum Length}
\label{sec:converge_vac}
As emitted electrons travel away from the graphene, periodic boundary conditions eventually cause unphysical interaction between the electrons emitted from either side of the material.
In particular, Coulombic repulsion across the periodic boundary artificially decelerates emitted electrons, potentially leading to suppressed emission and partial reabsorbtion by the material.
These artificial interactions are strongest for the higher energy emissions produced by faster protons, and sufficient vacuum is needed to keep them negligible for the duration of the few-fs simulations.

\begin{figure}[h]
\includegraphics{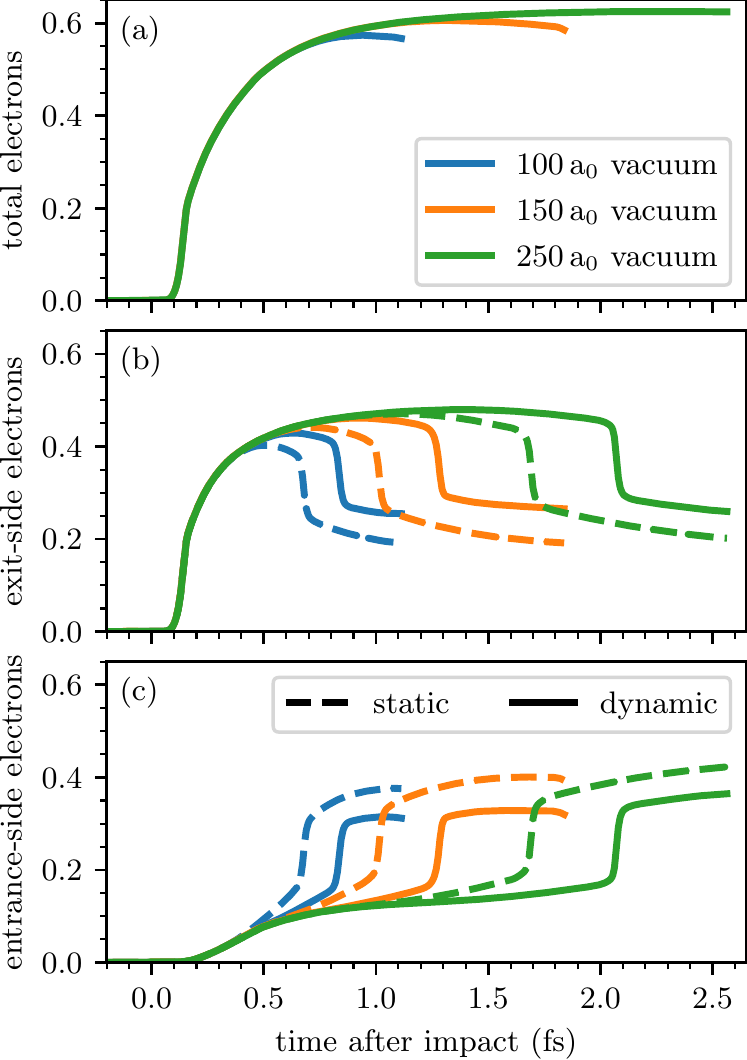}
\caption{
(a) The total number of electrons found in the vacuum, after a channeling proton with 1.8\,at.\,u.\ of velocity impacts graphene, is decomposed into (b) exit-side and (c) entrance-side regions using a static (dashed) and dynamic (solid) boundary between the two regions.
Results using three different vacuum lengths are compared, and the step-like features occur when the proton carries its captured electrons across the boundary between the exit-side and entrance-side regions.
}
\label{fig:vac_convergence}
\end{figure}

We assessed finite size effects arising from finite vacuum by repeating the case of a channeling proton with 1.79\,at.\,u.\ of velocity with three different vacuum lengths: 100\,a$_0$, 150\,a$_0$, and 250\,a$_0$ along the direction of proton motion.
Increasing the vacuum length both increases and delays the peak in total electron emission (see \mbox{Fig.\ \ref{fig:vac_convergence}}a) because more time passes before the exit-side and entrance-side distributions propagate toward the periodic boundary.
Also, increasing the vacuum length results in a slower decline of electron emission after the peak, indicating convergence towards a plateau without subsequent electron reabsorption.
Thus, we treat the maximum of the dynamic curve as the total number of electrons lost by the graphene (including both electron emission and electrons captured by the projectile). 
Predictions for this quantity differ by 3\% between the largest two vacuum lengths tested, so we conclude that the 150\,a$_0$ vacuum strikes an acceptable balance between accuracy and computational expense.

\section{Integration Boundaries}
\label{sec:dynamic_boundary}

\begin{figure}
\centering
\includegraphics{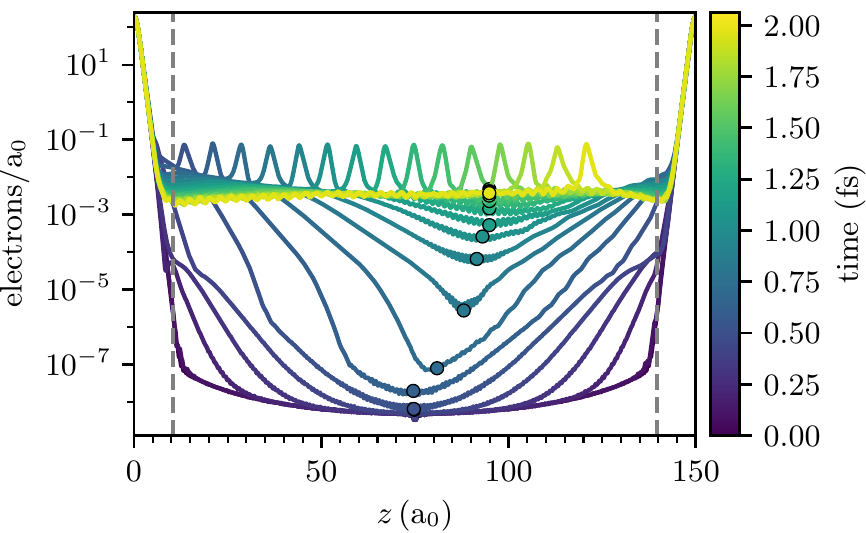}
\caption{\label{fig:dynamic_boundary}
Time-dependent distribution of electrons along the direction of proton motion ($z$) as a channeling proton with 1.8\,at.\,u.\ of velocity impacts graphene.
The electron density has been integrated over planes parallel to the graphene, which is located at $z=0$.
The smaller peaks which move to the right over time correspond to electrons captured by
the projectile.
Gray dashed lines indicate the boundaries between graphene and vacuum, while circles indicate the dynamic boundary between entrance-side and exit-side vacuum subregions.
}
\end{figure}

First, determining the total number of electrons in the vacuum requires definition of a material-vacuum boundary.
To eliminate contributions from the electron density extending from the graphene, we exclude the region within 10.5\,a$_0$ of the carbon atoms (see Fig. \ref{fig:dynamic_boundary}).
This cutoff gives less than $5\times 10^{-6}$ electrons in the vacuum for the ground-state initial condition.
We then decompose this total number of electrons into entrance-side and exit-side electrons with an additional boundary between the two vacuum subregions.

The simplest choice for the boundary between entrance-side and exit-side vacuum regions is a plane parallel to the graphene at the center of the vacuum.
However, with this choice, the entrance-side and exit-side electron emissions calculated using different vacuum sizes diverge much earlier than the corresponding total emissions (see Fig.\ \ref{fig:vac_convergence}), and an even larger vacuum would be needed to converge the number of electrons emitted from either side of the material individually.
This occurs because the finite vacuum and periodic boundary conditions obscure the distinction between electrons emitted from either side of the graphene as the time-dependent electron densities cross the vacuum center and eventually merge.

Instead, we allow the boundary between entrance-side and exit-side vacuum regions to shift off-center over time.
This dynamic, asymmetric treatment is motivated by the fact that exit-side electrons are generally more numerous (see Fig.\ 5 in the main text).
As this dynamic boundary, we take the plane parallel to the graphene where the average electron density achieves an interpolated minimum (see Fig.\ \ref{fig:dynamic_boundary}) as determined by quadratic fitting.
This choice delays the divergence between dynamical emissions calculated using different vacuum sizes and improves convergence of the exit-side and entrance-side yields with respect to vacuum size (see Fig.\ \ref{fig:vac_convergence}).

Similar to the above treatment of total electrons in the vacuum, we take the maximum of the corresponding dynamic curve as the number of exit-side electrons (including exit-side emitted electrons and electrons captured by the projectile).
Subtracting the number of exit-side electrons from the total number of electrons in the vacuum then gives a consistent number of entrance-side electrons.
By comparing results from the largest two vacuum sizes tested, we estimate that the number of exit-side and entrance-side electrons are converged to within 4\% with respect to vacuum size.

\section{Source of discrepancies in computational stopping results}

Zhao \emph{et al.}\ proposed that the differences in transferred energy for the centroid trajectory (see Fig.\ 1 in the main text), which are as large as 5\,eV, could arise from different pseudopotential types used in the different computational studies \cite{zhao:2015}.
However, we note that our results using HSCV norm-conserving pseudopotentials \cite{vanderbilt:1985} match more closely with the work of Ojanper\"{a} \emph{et al.}\ \cite{ojanpera:2014}, which used the projector augmented wave (PAW) formalism \cite{blochl:1994}, than with the work of Zhao \emph{et al.}\ \cite{zhao:2015}, which used norm-conserving Troullier-Martins pseudopotentials \cite{troullier:1991}.
While different pseudopotential cutoff radii may play a role (see Table \ref{tab:lit}), we note that the variation among computed results could also arise from finite size effects:
Zhao \emph{et al.}\ \cite{zhao:2015} used a smaller supercell than both Ojanper\"{a} \emph{et al.}\ \cite{ojanpera:2014} and the present study (see Table \ref{tab:lit}).
An earlier work on proton-irradiated graphene fragments \cite{bubin:2012} reported that energy transferred near the stopping power maximum is about 25\% higher for C$_{54}$H$_{18}$ fragments than for C$_{24}$H$_{12}$ fragments, a difference certainly large enough to help explain the 7\% discrepancy found here for graphene sheets.

\begin{table}[h]
\centering
\begin{tabular}{c|c|c|c}
& atoms & pseudopotential & cutoff radius \\\hline
Ojanper\"{a} \cite{ojanpera:2014} & 50 & PAW \cite{blochl:1994} & unknown \\
Zhao \cite{zhao:2015} & 24 & norm-conserving \cite{troullier:1991} & 1.45\,at.\,u.\\
this work & 112 & norm-conserving \cite{vanderbilt:1985} & 0.8\,at.\,u.
\end{tabular}
\caption{\label{tab:lit}
Different computational parameters used in this and earlier studies that investigate energy deposited by a proton impacting graphene along the centroid trajectory.
}
\end{table}

\section{Supplementary stopping results}

\begin{figure}
\centering
\includegraphics{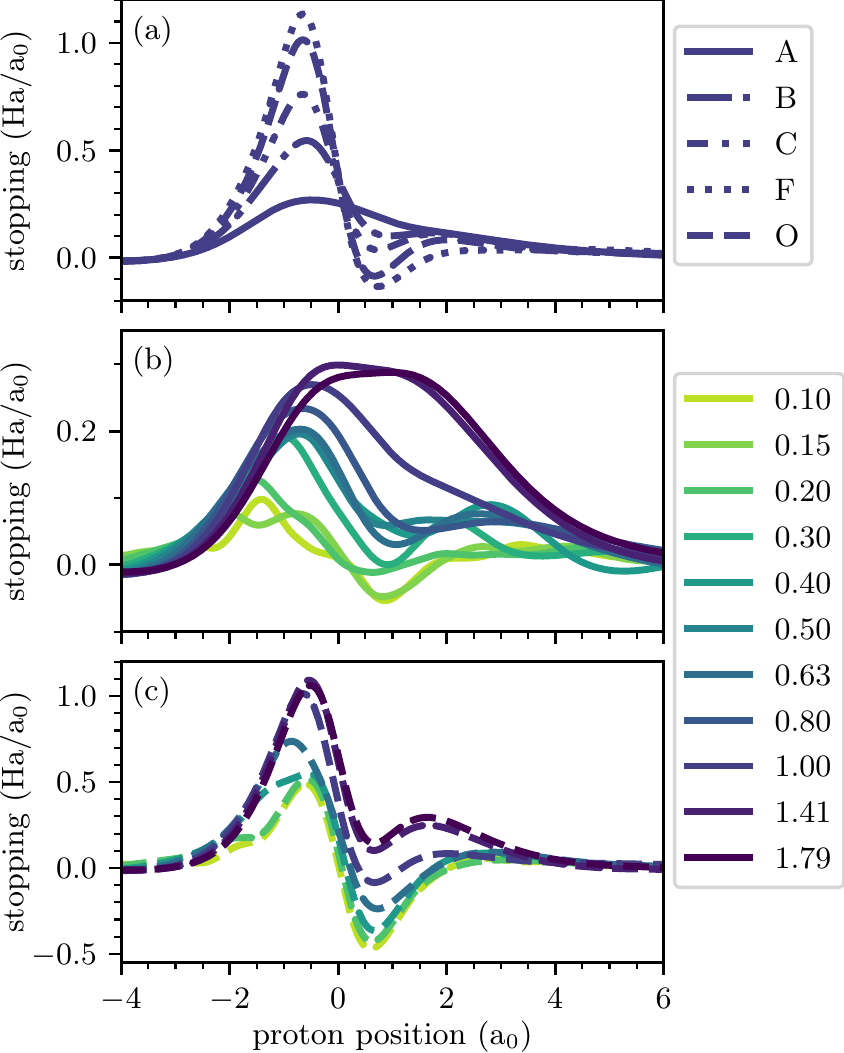}
\caption{Instantaneous stopping power as protons with (a) 1\,at.\,u.\ of velocity impact at different points (see Fig.\ 1a of the main text), (b) different velocities impact at point A (channeling trajectory), and (c) different velocities impact at point O (centroid trajectory).
}
\label{fig:TD_stop_panels}
\end{figure}

\begin{figure}
\includegraphics{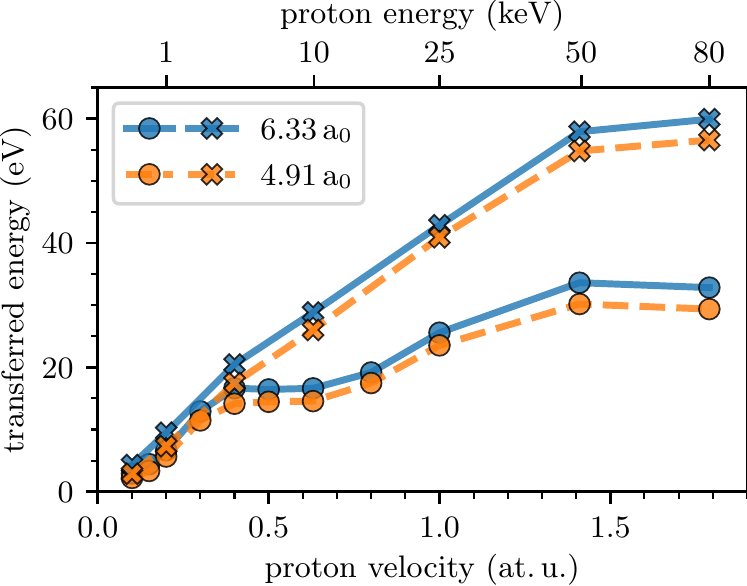}
\caption{Total energy deposited in graphene by energetic protons as determined using two different graphene thicknesses: the interlayer separation of 6.33\,a$_0$ in graphite and the dielectric thickness of 4.91\,a$_0$ \cite{rickhaus:2020}.
Circles (exes) indicate a channeling (centroid) trajectory (see Fig.\ 1a in the main text).
}
\label{fig:gr_thickness}
\end{figure}

Here we include additional data related to the energy deposition of protons in graphene.
First, Fig.\ \ref{fig:TD_stop_panels} shows the instantaneous stopping power across different proton trajectories and velocities.
For the total energy transfer presented in the main text, this data is integrated over a layer thickness of 6.33\,a$_0$, the interlayer separation in graphite.
Instead integrating over the dielectric thickness of 4.91\,a$_0$ reported in Ref.\ \onlinecite{rickhaus:2020} gives very similar results (see Fig.\ \ref{fig:gr_thickness}) and leaves the anomalous trends unchanged.

\section{Time-dependent Excited Electron Contributions}

\begin{figure}[h]
\centering
\includegraphics{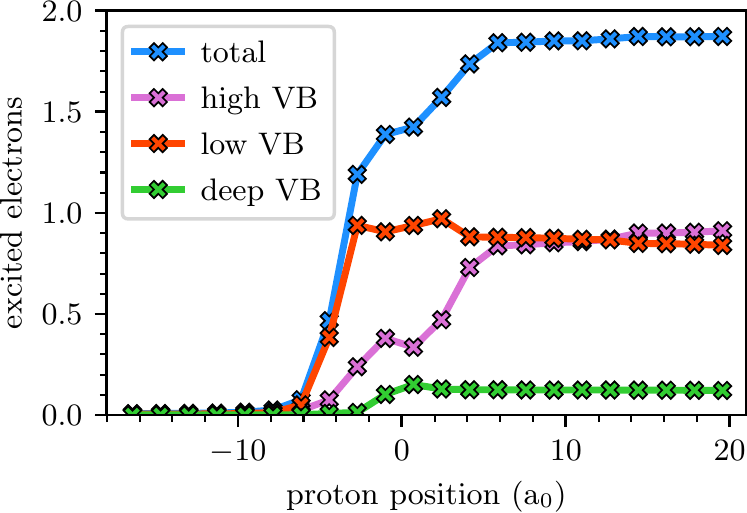}
\caption{Total number of excited electrons decomposed into contributions from different sets of valence states (see \mbox{Fig.\ 1a} in the main text) as a proton with 0.1\,at.\,u.\ of velocity traverses graphene along the centroid trajectory.}
\label{fig:eexcited_t}
\end{figure}

Fig.\ \ref{fig:eexcited_t} demonstrates that the excited electron contributions calculated according to Eq.\ (3) in the main text approach constant values after the proton moves sufficiently far from the graphene.
Thus, in Fig.\ 3 of the main text we simply report the number of excited electrons at the end of each simulation.

\section{Emission Dynamics}

\begin{figure}[h]
\centering
\includegraphics{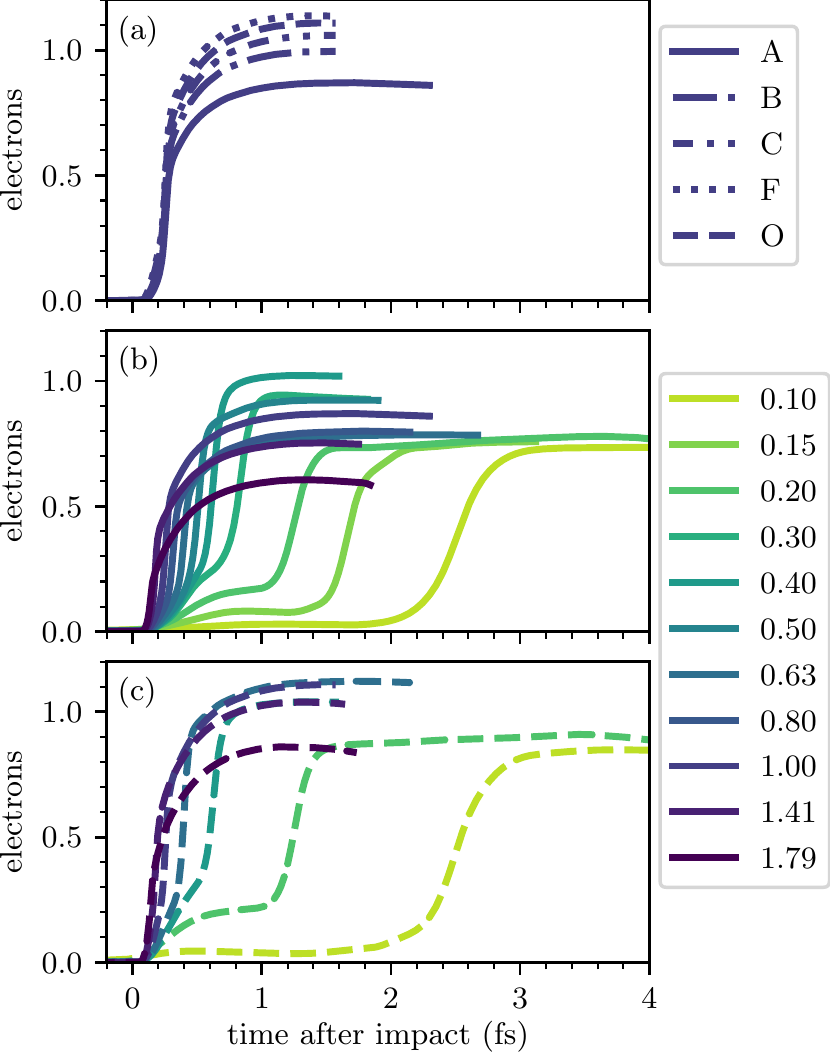}
\caption{Total number of electrons outside graphene after protons with (a) 1\,at.\,u.\ of velocity impact at different points (see Fig.\ 1a of the main text), (b) different velocities impact at point A (channeling trajectory), and (c) different velocities impact at point O (centroid trajectory).
Step-like features occur when slow protons exit into the vacuum region.
}
\label{fig:TD_SE_panels}
\end{figure}

Fig.\ \ref{fig:TD_SE_panels} provides information about the dynamics of electron emission across different proton trajectories and velocities.
Note that this data includes electrons captured by the proton, resulting in step-like features upon the projectile's exit into the vacuum region.
We attribute the slight decline in the total number of electrons in the vacuum to finite size effects (see \mbox{Fig.\ \ref{fig:vac_convergence}} and earlier discussion of vacuum length convergence).
Other than that, we observe no recapture of electrons by the graphene, which we would expect to appear as a significant drop in the total number of electrons in the vacuum within about 1\,fs of impact.

\section{Threshold emission models}
The binary elastic collision model of Ref.\ \onlinecite{baragiola:1979} predicts a threshold velocity $v_{\mathrm{th}}$ below which no electron emission occurs given by
\begin{equation}
v_{\mathrm{th}} = \frac{v_F}{2}\left(\left( 1+ \frac{\phi}{E_F} \right)^{1/2} -1\right),
\label{eq:vth_baragiola}
\end{equation}
where $v_F$ is the Fermi velocity, $E_F=\frac{1}{2}v_F^2$ is the Fermi energy, and $\phi$ is the work function.
For the case of graphene, Eq.\ \eqref{eq:vth_baragiola} can be evaluated using $v_F=1.1\times 10^6$\,m/s from experiment \cite{deacon:2007} or $0.89\times 10^6$\,m/s from DFT-LDA \cite{gillen:2010} and $\phi=4.56$\,eV from experiment \cite{yan:2012} or $4.37$\,eV from DFT-PBE \cite{leenaerts:2017}, giving an estimate of $v_{\mathrm{th}} = 0.13\,\--\,0.15$\,at.\,u.
Two other models can be adapted from the stopping power literature \cite{lim:2016,quashie:2016,correa:2018} by treating the work function as an energy gap:
\begin{equation}
\label{eq:vth_quashie}
v_{\mathrm{th}} = \frac{\phi}{2v_F} = 0.17\,\--\,0.19\,\mathrm{at.\,u.}
\end{equation}
and 
\begin{equation}
\label{eq:vth_lim}
v_{\mathrm{th}} = \frac{\lambda\phi}{h} = 0.16\,\--\,0.18\,\mathrm{at.\,u.},
\end{equation}
where $\lambda$ is the layer thickness (6.33\,a$_0$).


\bibliography{main.bib}

\end{document}